\documentclass[reprint,amsmath,amssymb,aps,prb,showpacs]{revtex4-1}

\usepackage{graphicx}
\usepackage[dvips]{color}

\begin{document}

\title{Wide-range optical studies on various single-walled carbon nanotubes: the origin of the low-energy gap}

\author{\'A. Pekker}
\email[]{pekkera@szfki.hu}

\author{K. Kamar\'as}

\affiliation{Research Institute for Solid State Physics and Optics, Hungarian Academy of Sciences, P.O. Box 49, Budapest, Hungary H-1525}

\date{\today}

\begin{abstract}
We present wide-range (3 meV - 6 eV) optical studies on freestanding transparent carbon nanotube films, made from nanotubes with different diameter distributions. In the far-infrared region, we found a low-energy gap in all samples investigated. By a detailed analysis we determined the average diameters of both the semiconducting and metallic species from the near infrared/visible features of the spectra. Having thus established the dependence of the gap
value on the mean diameter, we find that the frequency of the low energy gap is increasing with increasing curvature. Our results strongly support the explanation of the low-frequency feature as arising from a curvature-induced gap instead of effective medium effects. Comparing our results with other theoretical and experimental low-energy gap values, we find that optical measurements yield a systematically lower gap than tunneling spectroscopy and DFT calculations, the difference increasing with decreasing diameter. This difference can be assigned to electron-hole interactions.
\end{abstract}


\maketitle

\section{\label{sec:intro}Introduction}

Owing to the development of carbon nanotube-growing methods, a wide variety of single-walled nanotube samples is available at present. The most significant parameter of  these differently produced samples is their diameter distribution. All the important properties of nanotubes depend in some way on the diameter. The components of an ensemble can be identified by macroscopic characterization techniques which are usually optical methods. The most widely used techniques for this purpose are Raman spectroscopy,\cite{stranojacs03,stranonl03} photoluminescence spectroscopy,\cite{bachilosci02,oconnell02} and transmission spectroscopy. The first two are particularly suitable to determine the $(n,m)$ chiral indices of the constituting nanotubes. Transmission spectroscopy lacks the selectivity of the previous techniques and measures all nanotubes simultaneously; however, the resulting materials properties as transmission and refractive index have special importance in applications.

Here we present a detailed analysis of the optical properties of several single-walled carbon nanotube samples based on their wide-range transmission spectra. We concentrate on the low-frequency properties which have not been as widely investigated yet as the transitions between Van Hove singularities in the near infrared and visible region. We find a correlation between the low-frequency gap and the diameter. This correlation can be explained by a curvature-induced intrinsic gap\cite{kane97,kampfrath08} and does not agree with the predicted diameter dependence of effective-medium models.\cite{akima06,slepyan10} Our results, including the diameter dependence, agree very well with other optical studies on similar nanotube samples; however, a significant difference appears at low diameters between optical and both tunneling and density functional theory (DFT) data. We assign these differences to possible localized states even in these small gaps.

\section{\label{sec:exp}Experimental procedures}

The purpose of this paper is to determine the low-frequency optical properties of various carbon nanotubes and relate their low-frequency gap (if present) to the diameter of the tubes. For the accurate determination of optical functions, we use wide-range spectroscopy on self-supporting transparent films. The information about the gap is contained in the low-frequency end of the spectrum; in addition, the diameter distribution can be extracted from the structure in the near-infrared/visible range. This procedure has the advantage that both quantities are provided by the same experiment from exactly the same samples. The difficulties that arise stem partly from the overlap of interband transitions due to the distribution of diameters in the bulk samples, and partly from bundling which slightly changes the transition energies. To correct for these effects, we compare our results with Raman and photoluminescence data, taken on suspended tubes, from the literature.

Below, we show how we determined both the gap values and the diameter distributions and mean diameters of our samples. During the procedure, we also obtained the chiralities of the most abundant nanotube species in all the samples investigated which we present as supplemental material.\cite{epaps00}

\subsection{Nanotube samples}

We investigated and compared the optical spectra of seven different nanotube samples. All of them contained single-walled carbon nanotubes produced by different preparation techniques or modified in some way (Table~\ref{tab:table1}). Samples P2 and P3 were produced by arc discharge. Sample P2 was purified by oxidation in air,\cite{shi99} while P3 was treated with nitric acid. Samples Laser and Laser-H were made by laser ablation. The Laser sample was purified using different acids.\cite{rinzler98} Laser-H is similar to the Laser sample but annealed in order to remove the doping due to the purification process. HiPco tubes were produced by chemical vapor deposition (CVD) using high pressure carbon monoxide as carbon source. CoMoCat-CG was produced by CVD on a cobalt-molybdenum catalyst. The SG variant of the CoMoCat tubes was enriched with semiconducting tubes, with more than 50 per cent (6,5) tubes.

\begingroup
\squeezetable
\begin{table*}
\caption{\label{tab:table1}
Different nanotube samples used in the comparison. The indicated average diameters are based on the optical measurements, see text.}
\begin{ruledtabular}
\begin{tabular}{lllc}
\textrm{Sample}& \textrm{Company}& \textrm{Note}&
\textrm{Average diameter (nm)}\\
\colrule
P2 & Carbon Solutions & $O_2$ purified arc-discharge tubes& 1.42\\
P3 & Carbon Solutions & acid treated arc-discharge tubes& 1.42\\
Laser & Rice University & acid treated laser ablation tubes& 1.25\\
Laser-H & Rice University & annealed Laser sample (400$^{\circ}$C 12h) & 1.25\\
HiPCO & CNI Nanotechnologies & CVD tubes& 1.08\\
CoMoCat CG & Southwest Nanotechnologies & CVD using Co-Mo catalyst& 0.90\\
CoMoCat SG & Southwest Nanotechnologies & CoMoCat sample enriched with semiconducting tubes & 0.76\\
\end{tabular}
\end{ruledtabular}
\end{table*}
\endgroup

\subsection{Wide-range spectroscopy}

Self-supporting thin films for wide-range transmission measurements were produced by vacuum filtration, following the recipes given in Ref. \onlinecite{wu04}. Using the freestanding samples we can perform the measurements on the same sample in a wide frequency range. This method circumvents the inconveniences due to the limited transmission window of substrates. Aqueous nanotube suspensions were produced using Triton-X as surfactant. The suspension was left for sedimentation, and the supernatant was filtered through an acetone-soluble filter. The filter was dissolved in acetone and the resulting nanotube thin film was stretched over a hole on a graphite substrate. Mild annealing was applied to remove the solvent from the sample.

Different instruments were used to measure wide-frequency (20-55000 cm$^{-1}$) transmission spectra: a Bruker IFS 66/v vacuum Fourier transform infrared (FTIR) spectrometer in the far-infrared (FIR) and mid-infrared (MIR) region, a Bruker Tensor 37 FTIR spectrometer in the near infrared (NIR), and a Jasco v550 grating spectrometer in the visible and ultraviolet (UV). We used a standard transmission arrangement with normal incidence. Spectral resolution was typically 2~cm$^{-1}$ in the FIR-NIR range, and 1~nm in the visible-ultraviolet range.

\subsection{Kramers-Kronig calculation}

\begin{figure}[ht!]
\includegraphics[width=8 cm]{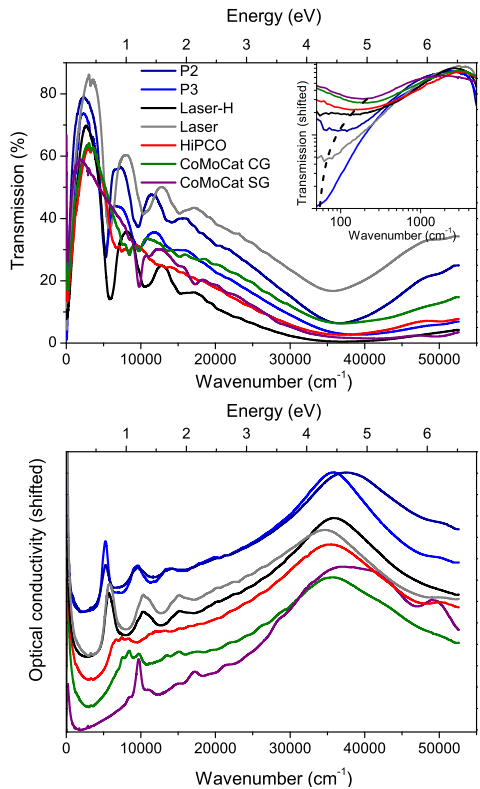}
\caption{\label{fig:T_OC}(Color online) The measured wide range transmission spectra of the different nanotubes, and the calculated optical conductivity using the Kramers - Kronig relations and model considerations (for details, see text). Curves have been shifted along the y axis for clarity. The inset, on a log-log scale, indicates that the minimum in transmission increases in frequency with increasing mean diameter.}
\end{figure}

We calculated the optical conductivity from the transmission data using the Kramers-Kronig equations.\cite{borondics06,pekker06} Optical functions derived from samples of different thickness are thus directly comparable (Fig.~\ref{fig:T_OC}). In order to perform this calculation we have to use a model for our samples. We consider the self-supporting sample as a homogeneous layer with finite thickness and parallel surfaces. The optical properties of such a layer are determined from the measured transmission by the Fresnel equations. The transmission coefficient $t$ is:

\begin{equation}\label{Fresnel}
    t=\sqrt{T}e^{i\phi}=\frac{4N}{(N+1)^2e^{-i\delta}-(N-1)^2e^{i\delta}},
\end{equation}
where
\begin{equation*}
\quad\delta=\frac{\nu^* Nd}{c},
\end{equation*}
$\nu^*$ is the wavenumber of the light, $c$ the speed of light, N the complex refractive index of the material and $d$ the thickness of the sample. The measured quantity is the transmittance (T), and the corresponding phase change ($\phi$) is related to T by the Kramers-Kronig equation:
\begin{equation}\label{KK}
\phi(\nu_0^*)=2\pi d\nu_0^*-\frac{2\nu_0^*}{\pi}\int_{0}^{+\infty}\frac{\ln{(\sqrt{T(\nu^*)}/\sqrt{T(\nu_0^*)})}}{\nu^{*2}-\nu_0^{*2}}d\nu^*.
\end{equation}

Formally, the phase-shift integral requires knowledge of the
transmittance at all frequencies. The measurements on freestanding films allow us to
obtain accurate transmission data over three orders of magnitude in frequency; nevertheless, extrapolations are needed to
complete the transform above and below the range of the available measurements. Since we are mostly interested in the low-frequency behavior, the choice of extrapolation towards zero frequency is more critical. The conventional low-frequency extrapolation for metals is
 $T(\omega) = T(0) + A\omega^2$, where $A$ is
a constant determined by the transmittance at the lowest frequency
measured in the experiment and $T(0)$ is the extrapolated behavior to
zero frequency, determined  by the dc conductivity. For semiconductors, the transmittance is continued smoothly towards zero frequency. The high-frequency
extrapolation uses $T = 1 - C\omega^{-n}$ with $n\approx 1$ and $C$ chosen to give a smooth
connection to the high-frequency transmittance curve. In our calculations, the low-frequency part was insensitive to the details of both low- and high-frequency extrapolations.

\begin{figure*}
\includegraphics[width=16cm]{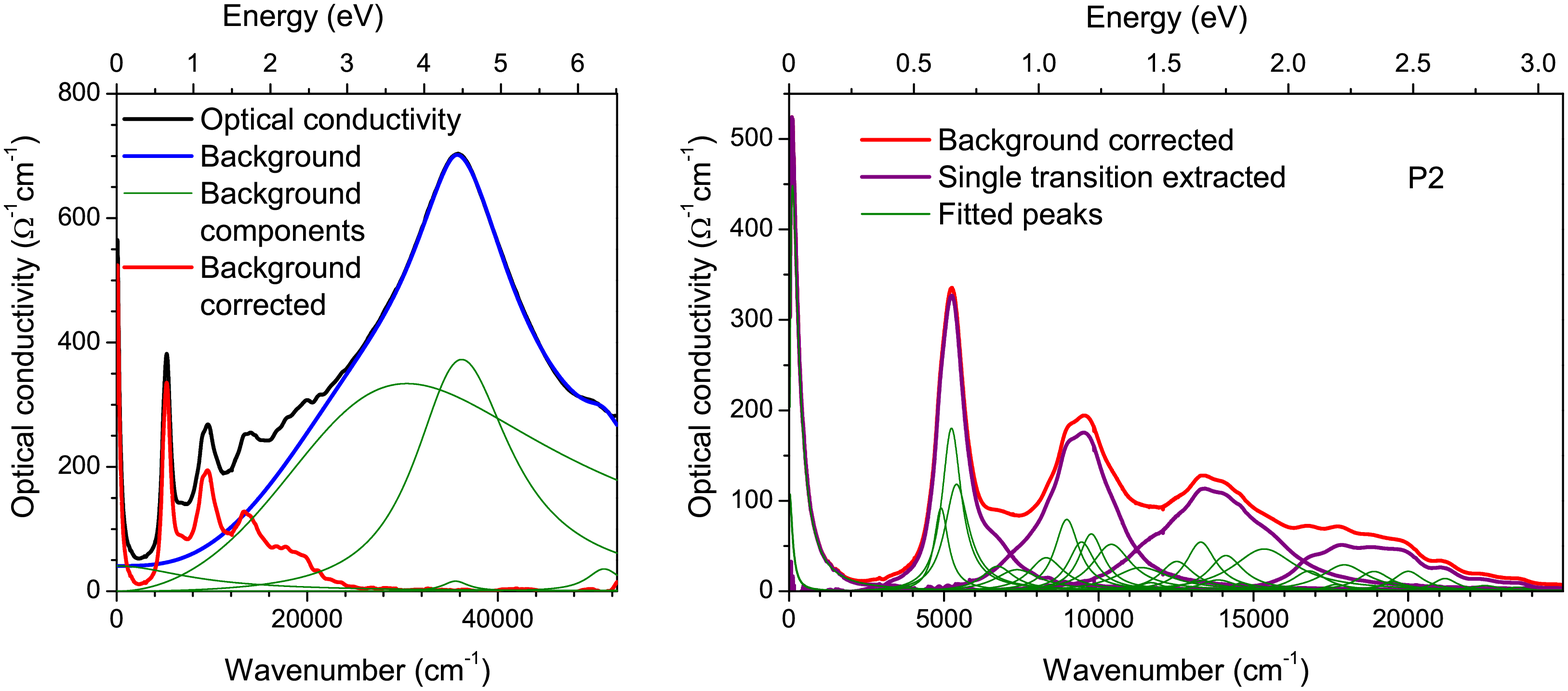}
\caption{\label{fig:fit}(Color online) Baseline correction of the optical conductivity of the P2 sample using a Drude-Lorentz fit. The peaks were assigned either to the spectral features or to the background (the $\pi$-$\pi^*$ transition and a constant from carbonaceous impurities at low frequencies is considered as background). After the background subtraction the spectra can be further analyzed and a single set of peaks can be selectively extracted (see text for explanation).}
\end{figure*}

 Our simple single-layer model does not take into account that the nanotube thin film is a porous structure, a random network composed of entangled nanotube bundles; consequently, the optical path length differs from the actual thickness. To eliminate the effects of different morphologies, the optical conductivity spectra were scaled to the $\pi$-$\pi^*$ transition by varying the thickness parameter during the Kramers-Kronig calculation. Approximate thickness values were obtained by atomic force microscopy (Veeco CP-II). The thickness of our films varied between 90 and 250 nm; they can be regarded as homogeneous for light transmission, with few unobstructed paths through the network.\cite{wu04} Due to the scaling, the quantitative information is restricted to the peak positions, the intensities can be used only for qualitative comparison between different nanotube samples.

\subsection{Drude-Lorentz fit and baseline correction}

For further analysis we fitted the optical conductivity by the Drude-Lorentz model:
\begin{equation}\label{DL}
\sigma_1 = {\epsilon_0 \left[{\frac{\omega_{p,D}^2\gamma_D}{\omega^2 + \gamma_D^2}}  +  \sum_i\frac{\omega_{p,i}^2\gamma_i\omega^2}{(\omega_{c,i}^2 - \omega^2)^2 + \gamma_i^2 \omega^2} \right]},
\end{equation}
where $\epsilon_0$ is the vacuum dielectric constant, $\omega_{p,D}$ and $\gamma_D$ the plasma frequency and width of the free-carrier (Drude) contribution, and $\omega_{c,i}, \omega_{p,i}$ and $\gamma_i$ the center frequency, generalized plasma frequency and width of the Lorentz oscillators corresponding to transitions of bound carriers. (The generalized plasma frequency for Lorentzians is the measure of the oscillator strength.)

The result of these fits is an unusually large number of oscillators, since the samples consist of several types of nanotubes with transition energies depending on their chirality.\cite{mintmire98} (Despite their large number, these features cannot be
related directly to single nanotube species. Especially in the case of large average diameter samples these peaks originate from many individual nanotubes, but due to their similar transition energies they cannot be decomposed further.)
For the transitions, we use the notation of Ref. \onlinecite{itkis02}: $M_{00}$ for the intraband/small gap transition of formally metallic tubes, $S_{11}, S_{22}, M_{11}, M_{22}$... the interband transitions of semiconducting and metallic tubes, respectively, in order of increasing energy. The full set of fit parameters is presented in Ref. \onlinecite{epaps00}.

\begin{figure}
\includegraphics[width=8 cm]{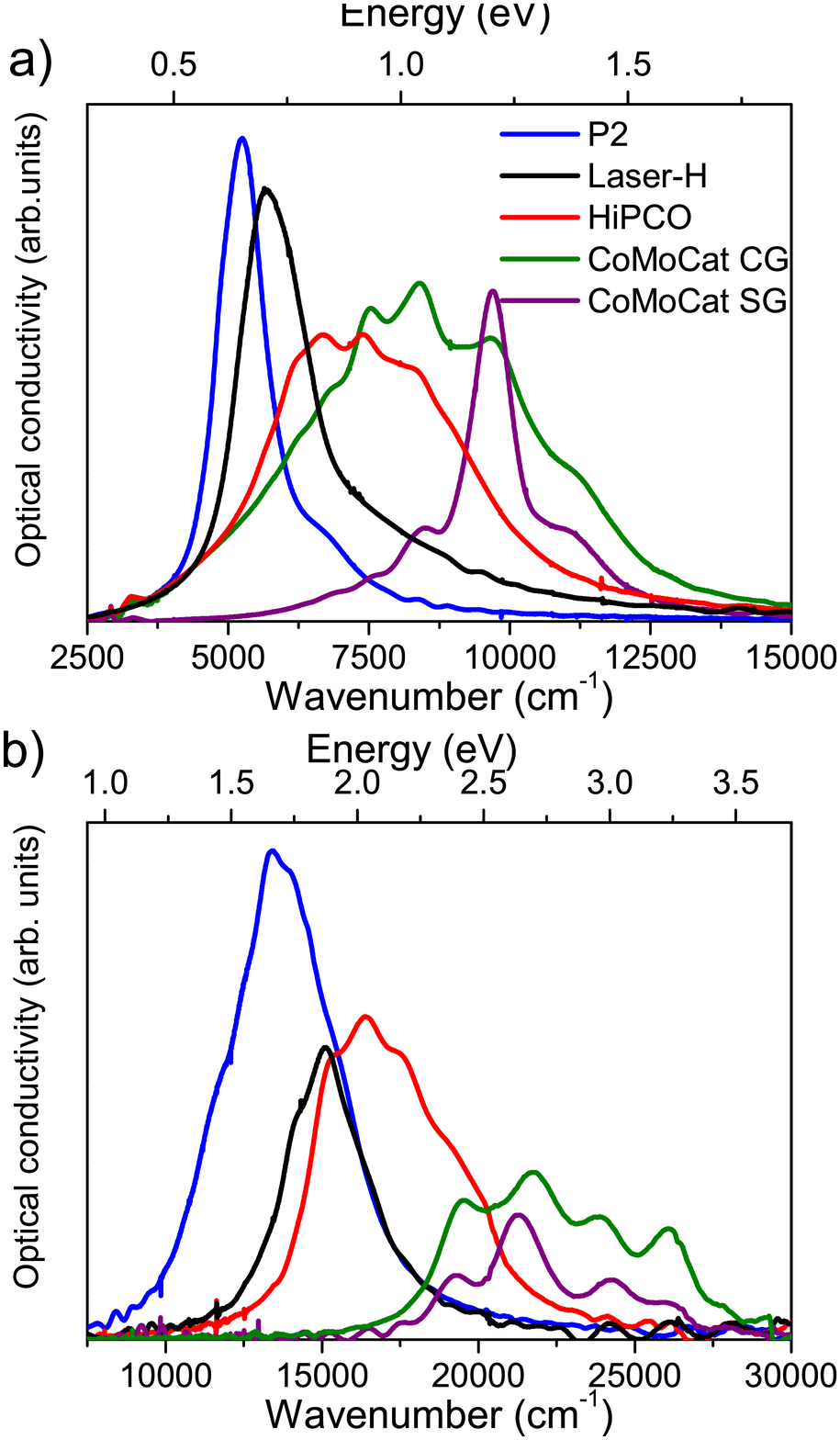}
\caption{\label{fig:S11M11}(Color online) a) Comparison of $S_{11}$ transition peaks of our samples. The samples possess different diameter distribution which appear in their optical spectra. b) The $M_{11}$ peaks of the same samples.}
\end{figure}

When assigning a specific group of peaks to one type of transition, we have to distinguish between transitions belonging to the given group and all others, the latter considered background. In the first step, we have to distinguish the spectral information belonging to the one-dimensional structure of the tubes from other contributions, mainly the $\pi$-$\pi^*$ transition of the full $\pi$-electron system, and from those caused by other carbonaceous material present in almost all nanotube ensembles. (Although the samples used represent some of the highest quality commercial nanotube products, the problem cannot be eliminated completely and has to be considered when evaluating intrinsic properties.\cite{haddon04}) The former can be modeled by a few Lorentzians,\cite{murakami05prl} and the latter is considered a constant in the low-frequency regime.\cite{itkis03} This constant was taken into account as a weak and broad Drude oscillator. We treat the sum of these two effects
as background and subtract it to emphasize the free-carrier contribution and the electronic transitions of the nanotubes themselves (Fig.~\ref{fig:fit}).

In case of laser ablation and arc-discharge tubes the corrected spectrum contains more or less separated groups of peaks which can be assigned easily to the different transitions (in the sequence $M_{00}$, $S_{11}$, $S_{22}$, $M_{11}$). In the spectra of small diameter tubes like HiPco and CoMoCat the $S_{22}$ transitions of lower diameter nanotubes overlap with the $M_{11}$ transitions of the higher diameter ones. Therefore the assignment is somewhat ambiguous. It can be improved, however, utilizing the results of previous resonance Raman studies on suspended nanotubes.\cite{stranojacs03,stranonl03} These experimental investigations provide a database of electronic transition energies by nanotube species. Comparing the center frequencies of the fitted oscillators to these values helps us separate the $S_{22}$ and $M_{11}$ peaks. With the association of the oscillators with different transitions, the spectrum can be further optimized for analysis. In order to extract as much of the original information as possible related to one specific set of peaks ($S_{11}$, $S_{22}$, etc...), the Lorentzians assigned to other transitions are considered as background and subtracted. This procedure leads to spectra as depicted in Fig.~\ref{fig:S11M11}: we preserved the original data in the region of interest, containing all the small features which otherwise would have been lost.

\section{\label{sec:results}Results and analysis}

\subsection{Determination of the effect of bundling on the transition energies}

\begin{figure}
\includegraphics[width=8.5 cm]{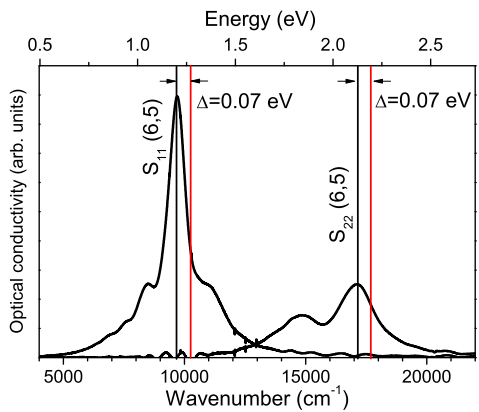}
\caption{\label{fig:delta}(Color online) The determination of the correction due to bundling. In the optical conductivity spectrum of the enriched CoMoCat sample the contribution of the (6,5) tube can be easily identified. The red lines show the transition energies of the individual (6,5) nanotube defined by photoluminescence measurements.\cite{bachilosci02} The same shift can be applied in both frequency ranges.}
\end{figure}

The optical behavior of the nanotubes is determined by their diameter. It is possible to deduct the diameter distribution of the samples from their optical spectra,\cite{liu02} but chiral index assignment cannot be performed based on transmission data alone. The transition energies of the different $(n,m)$ species can be determined from Raman \cite{stranojacs03,stranonl03} and photoluminescence \cite{bachilosci02,oconnell02} measurements on suspended tubes in solution. Previous studies show that bundling and other environmental effects produce a frequency shift between bulk and suspended tubes, which has to be obtained experimentally. \cite{fantini04,tan08,li09,oconnell04,wang06}. O'Connell \emph{et al.}\cite{oconnell04} determined this shift for various nanotube species using resonance Raman spectroscopy: the measured shifts show no correlation with diameter or chiral angle. They also found that the same shift can be applied for the different transitions ($S_{11}$, $S_{22}$). In our case we determined this shift using the CoMoCat SG sample. This sample is enriched with $(6,5)$ type semiconducting tubes. In the optical spectrum we can easily identify the transition peaks of the $(6,5)$ type tubes (Fig.~\ref{fig:delta}). Comparing the center of these peaks to the values obtained by measurements on individual tubes,\cite{bachilosci02} we can determine the shift due to the environment ($\Delta=0.07$ eV) which is in good agrement with other experimental results.\cite{oconnell04,wang06}.

We used this value to correct not only the CoMoCat SG but all other nanotube spectra. This generalization is supported by the findings of O'Connell\cite{oconnell04} and based on the presumed similarity of the environmental effects in case of different nanotube samples. The main origin of the shift is the dielectric screening due to the neighboring nanotubes.\cite{wang06} In the bundle the dielectric environment is supposed to be independent of the diameter of the nanotubes, change only with the size of the bundle, and probably saturate as the size increases. Our solid samples presumably contain large bundles which means we are already in the saturation range and a constant can be applied. Applying this correction to the spectra, the transition peaks become directly comparable to the Raman and photoluminescence measurements mentioned above.

\subsection{Determination of the diameter distribution}

We used the first transitions of the semiconducting and the metallic nanotubes, respectively, to define the diameter distribution of our materials. For the diameter determination we use the imaginary part of the dielectric function $\epsilon_2=\sigma_1/\omega$, where $\sigma_1$ is the optical conductivity. (Maxima in the joint density of states occur at maxima in $\epsilon_2$;\cite{reich04} these differ slightly from the peaks in $\sigma$ due to the factor $\omega$.) The most abundant nanotube types are determined using the first and third quantiles ($Q_1$, $Q_3$) of the background corrected peaks (Fig.~\ref{fig:S11M11}). Comparing the energy range defined by $Q_1$ and $Q_3$ to the transition peaks of individual nanotube species\cite{stranojacs03,stranonl03} we can determine the composition of our samples. In the case of the CoMoCat samples the $M_{11}$ peaks are in the visible region and merged into the $\pi - \pi^*$ background. In the spectrum, only the contributions from the most abundant metallic nanotubes are detectable, thus we do not have to use the quantiles method to determine the most probable nanotube species. In this case the energy ranges related to the peaks were determined by the parameters of the assigned Lorentzians: $[\omega_{c,L}-\frac{\Gamma_L}{2},\omega_{c,H}+\frac{\Gamma_H}{2}]$, where $\omega_{c,L}$ and $\Gamma_L$ are the center and the width of the Lorentzian with the lowest energy, and $\omega_{c,H}$ and $\Gamma_H$ are the same parameters related to the Lorentzian with the highest energy. The determined wavenumber ranges were converted to energy and corrected by the above-mentioned 0.07 eV shift. Comparing these energy ranges to the transition peaks of individual nanotube species\cite{stranojacs03,stranonl03} we can determine the composition of our samples. To characterize the samples, we determined the average diameters of the semiconducting and metallic fraction and for the whole ensemble, respectively. Table~\ref{tab:diameters} shows the result of this procedure. The details can be found in Ref. \onlinecite{epaps00}.

\begin{table}
\caption {Average diameters of the semiconducting,
metallic and non-armchair metallic fractions and the
average diameter of the whole ensemble in the case of
different samples (in nm). About the purpose of
non-armchair average diameter, see Section \ref{sec:disc}.}

\label{tab:diameters}
\centering
\vspace{10px}
\begin{tabular}{c|c|c|c|c}
& Semiconducting & Metallic & Non-armchair & Overall\\
&                &          & metallic     &         \\
\hline\hline
        P2 &   1.43 &  1.41 & 1.42 & 1.42 \\

   Laser-H &   1.25 &  1.25 & 1.25 & 1.25 \\

     HiPco &   1.07 &  1.11 & 1.12 & 1.08 \\

CoMoCat CG &   0.97 &  0.77 &  0.79 & 0.90 \\

CoMoCat SG &   0.74 &  0.77 &  0.79 & 0.76 \\
\hline\hline
\end{tabular}

\end{table}

\begin{figure}
\includegraphics[width=8 cm]{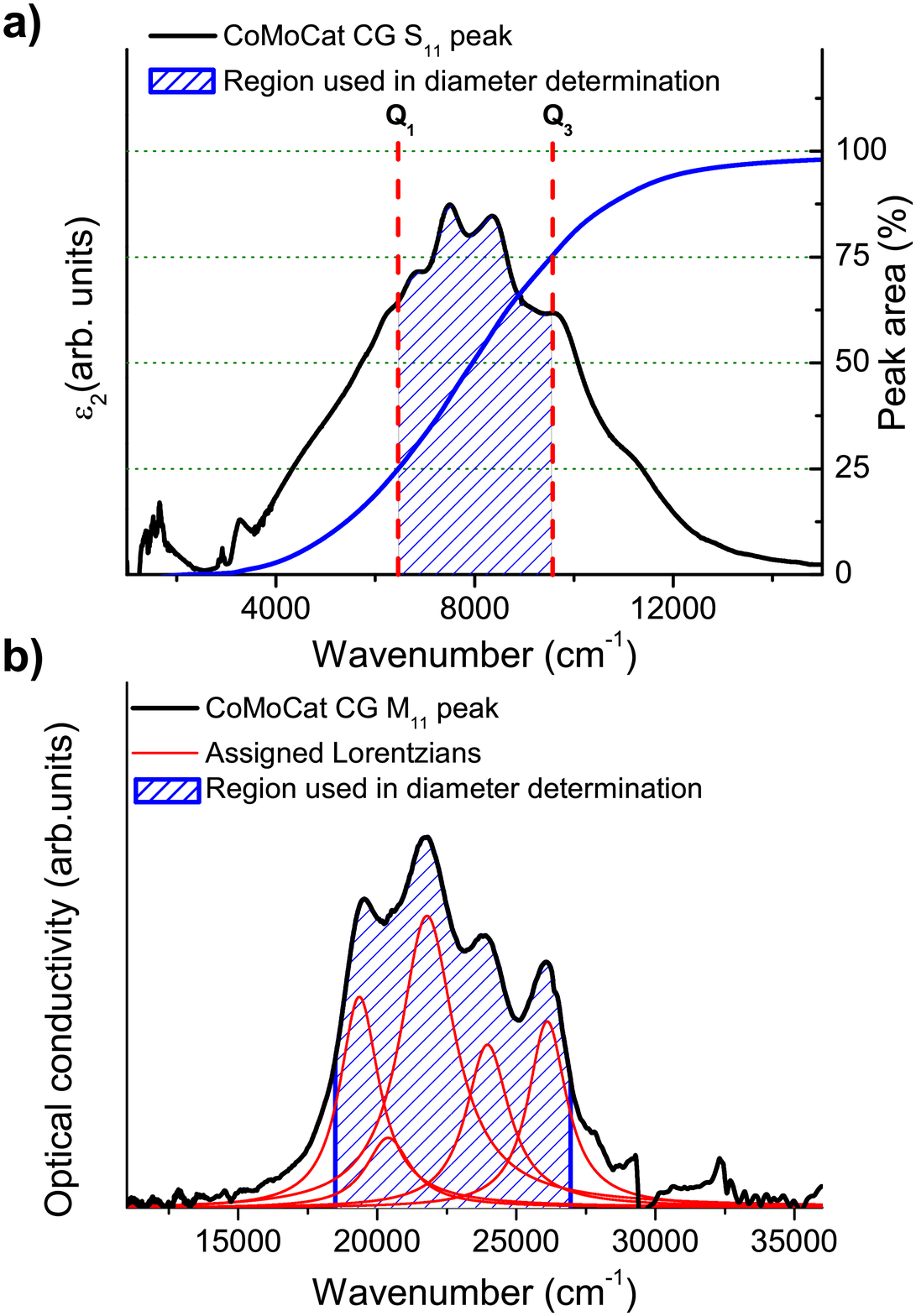}
\caption{\label{fig:diameter} (Color online) The average diameter determination for the CoMoCat CG sample. First we determine those regions of the $S_{11}$ and $M_{11}$ peaks which are related to the most abundant nanotube species. a) Except the $M_{11}$ peak of the CoMoCat samples, the energy range for diameter determination was calculated using the peak area. The blue curve is the integrated peak intensity as a function of wavenumber. The first ($Q_1$) and third ($Q_3$) quantiles refer to those wavenumbers where the area equals 25\% and 75\% of the whole peak area, respectively. b) In the case of CoMoCat samples, the $M_{11}$ peaks are merged into the $\pi-\pi^*$ background, therefore only the signatures of the most abundant nanotubes are detectable. In this case we can use directly the parameters of the assigned Lorentzians to determine the diameter distribution. See text for more explanation.}
\end{figure}

\section{\label{sec:disc}Discussion}

\begin{figure*}
\includegraphics[width=16 cm]{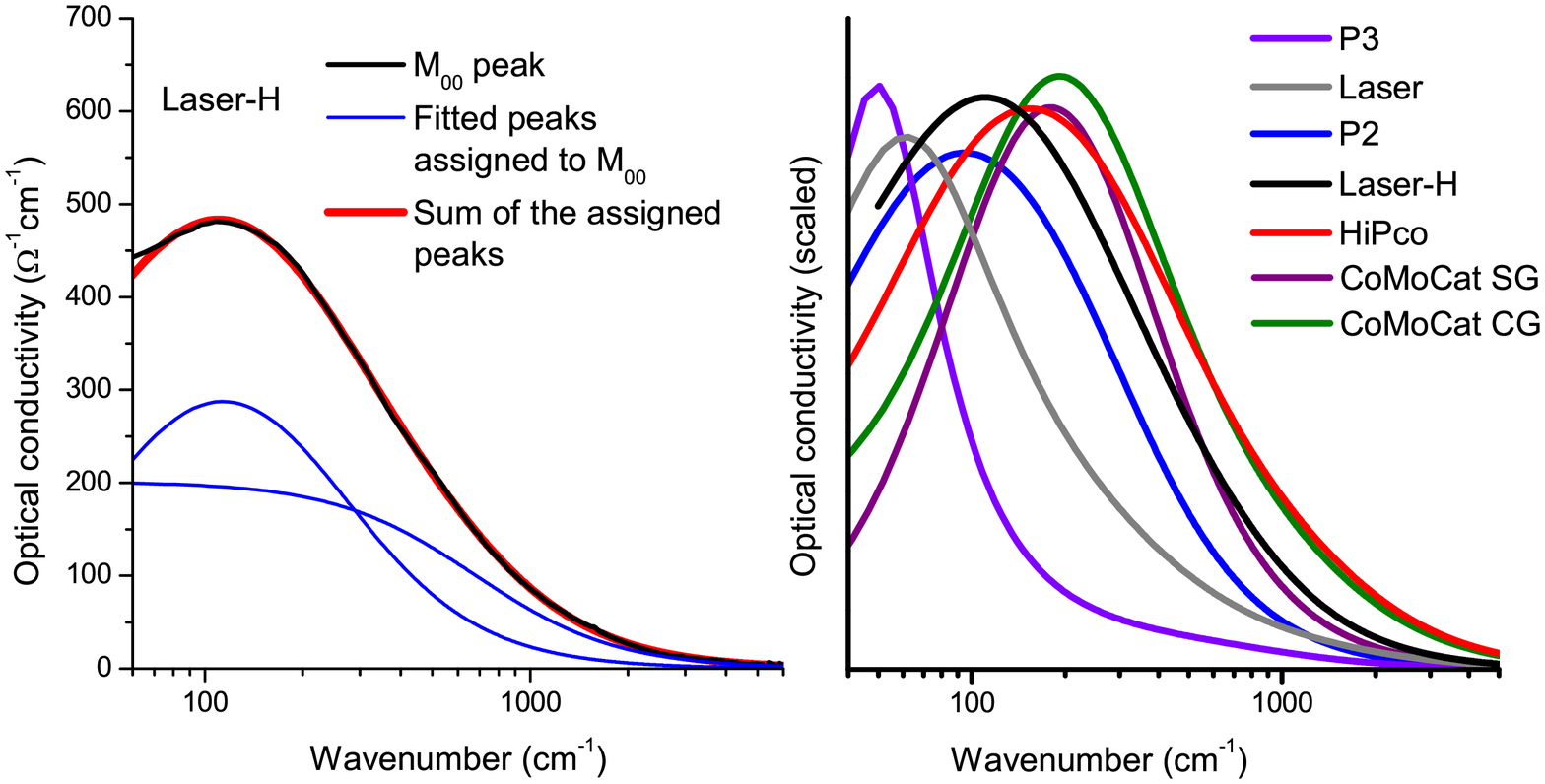}
\caption{\label{fig:Eg} (Color online) Left panel: The extracted low frequency peak (M$_{00}$) of the Laser-H sample with the fitted oscillators and their sum. Note the logarithmic frequency scale. The low frequency gap was defined using the fitting parameters from Ref. \onlinecite{epaps00}. Right panel: The fitted M$_{00}$ curves of all samples, clearly indicating the variation in the gap energy.}
\end{figure*}

The electronic structure of carbon nanotubes is determined by their $(n,m)$ wrapping indices. In their classic paper on the electronic structure of carbon nanotubes, Hamada, Sawada and Oshiyama\cite{hamada92} predicted the $(3n,0)$ zigzag nanotubes to possess a narrow gap of the order of 10 meV, decreasing with increasing diameter, in contrast to truly metallic armchair $(n,n)$ tubes. These calculations have been extended to all tubes with diameters below 1.5 nm by Kane and Mele,\cite{kane97} with the result that except the armchair nanotubes, all others satisfying the $n\equiv m$(mod 3) condition develop a gap below 0.1 eV. The mechanism behind the inhibition of electric conductivity is the $\pi$-orbital misalignment\cite{niyogi02} increasing with increasing curvature.

On the experimental side, tunneling spectroscopy on individual nanotubes\cite{ouyang01} confirmed the presence of a low-energy gap. Low-frequency peaks have been reported several times in the optical conductivity or optical density of macroscopic nanotube samples,\cite{ugawa99,itkis02,borondics06,akima06} but their interpretation is not uniform. Part of the controversy stems from the evaluation procedures varying with the measurement method.

Strictly speaking, transitions through a gap cause a peak in the imaginary part of the dielectric function $\epsilon_2$ ($\epsilon_2 = \sigma_1/\omega$) at the gap value. This quantity cannot be measured directly, but is determined by Kramers-Kronig transformation from wide-range reflectivity or transmission of neat samples.\cite{wooten72,dresselgruner02} Power absorption is proportional to the imaginary part of the refractive index and contains contributions from both real and imaginary parts of $\sigma$; moreover, the optical density derived from the transmission as $-log T$ differs from the true absorption function because of corrections due to reflectance at the interfaces. Whether or not these factors can be neglected during the analysis depends on their exact values. For carbon nanotubes in the far-infrared region, the difference between absorption, optical conductivity and optical density can be significant.\cite{pekker06} Nevertheless, optical density is often used for comparison of samples, especially thin layers on a substrate, because the transmission measurement at normal incidence (using the substrate as reference) cancels the substrate contribution.

In a composite material, even the overall optical conductivity  can differ from that of the ingredients. Effective-medium theory predicts that small metallic particles in a dielectric medium will develop a finite-frequency peak in the conductivity of the composite. Elaborating on the effective medium theory, Slepyan et al.\cite{slepyan10} cite the finite length of the nanotubes as the crucial factor behind shifting the conductivity maximum of metallic nanotubes from zero to finite frequency.

\begin{figure}

\includegraphics[width=8 cm]{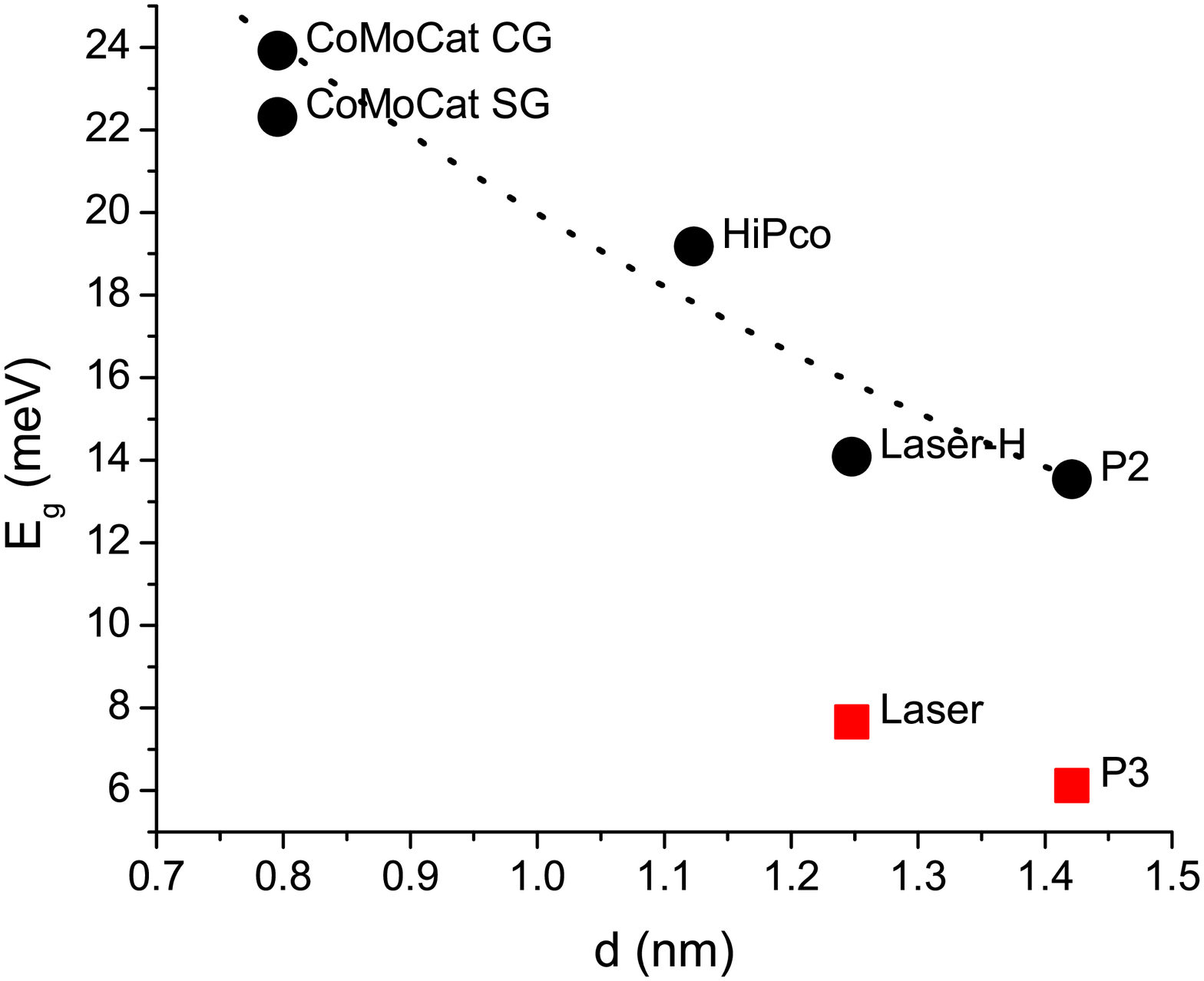}
\caption{\label{fig:EgE11} (Color online) Low-frequency gap position
versus non-armchair metallic average diameter for all samples measured.Black dots correspond to undoped samples and red squares to doped samples. The dotted line through the undoped data is a guide to the eye.}
\end{figure}

Our method to determine the low-energy gap is illustrated in Fig.~\ref{fig:Eg}. The right panel shows the low-frequency behavior in more detail. The gap energies $E_g$ were determined as the center frequency of the lowest frequency Lorentzian in the Drude-Lorentz fit (Fig.~\ref{fig:Eg}, details can be found in Ref. \onlinecite{epaps00}). It is obvious from the figure that all samples show a low-energy gap which increases with decreasing average diameter. In Fig.~\ref{fig:EgE11} we present our gap values as a function of the non-armchair metallic mean diameter from Table~\ref{tab:diameters}. In the following, we will put our results in perspective, based on previous knowledge summarized above, and compare them to other far infrared/terahertz experiments.

Itkis et al.\cite{itkis02} published a comprehensive study of optical density on spray-coated films of three types of nanotubes,
whose properties are close to some of the samples reported in this paper (arc-produced, laser and HiPco). All three samples exhibit a far-infrared peak in the optical density, its frequency increasing with decreasing mean diameter of the sample. Our optical conductivity data support their conclusions of the low-frequency gap causing the peaks.

A strong experimental proof for the low-frequency gap is the study by Kampfrath et al.\cite{kampfrath08} who examined the behavior of the far-infrared absorption on photoexcitation by a short visible laser pulse. Their model, based on an ensemble of two-level systems with a variation in the chemical potential, explains the observed spectrum and even its weak temperature dependence.\cite{borondics06}

Akima et al.\cite{akima06} have studied several samples, a "true" composite material (0.5 weight percent SWNT in polyethylene) and bulk nanotubes, sprayed on a silicon substrate. The composite exhibited a strong optical density peak in the far-infrared region, which they attributed to the Drude absorption of small metallic nanotube particles, shifted in frequency by effective-medium effects. They generalize this result to concentrated nanotube networks, although it is obvious even from their data that in a more concentrated sample, the peak appears at lower frequency. (They explain the latter as due to morphology and anisotropy.) We agree with Kampfrath et al.\cite{kampfrath08} that neat nanotube networks can be considered uniform at far-infrared frequencies, but at low concentrations isolated nanotube clusters can behave as metals in a dielectric.

The data in Fig.~\ref{fig:Eg} cannot be explained by the model of Slepyan et al.\cite{slepyan10} either, since that model predicts a very weak diameter dependence. They could be reconciled if the length distribution were correlated with the mean diameter, which, however, is highly improbable. The aspect ratio does not change considerably for nanotubes a few micrometers long, in the diameter range between 0.8 and 1.5 nm.

Adapting now the explanation of the peaks in the optical conductivity assigned to the curvature-induced secondary gap, we examine its diameter dependence and compare it to the values determined by other methods.

The low-energy gap occurs in all samples examined in the present study. At first glance the experimental values are randomly distributed in Fig.~\ref{fig:EgE11}, but if we make a distinction between the data of the modified (red squares) and the unmodified (black dots) samples, the latter show clear diameter dependence. The curvature-induced gap ($E_g$) was estimated to depend on \emph{d} as $1/d^2$ by both theoretical\cite{kleiner01,kleiner01a} and experimental\cite{ouyang01} studies, but according to density functional theory (DFT) calculations\cite{zolyomi04} an additional $1/d^4$ term improves the fits considerably. Although the tendency is clear, we cannot establish a quantitative connection between the diameter and the gap energy due to the averaged nature of the determined values. The gap which appears as a peak in the low frequency range of the optical conductivity spectrum related to the whole ensemble of the metallic nanotubes thus cannot be connected to a specific diameter or chirality. However, the gap value clearly increases with decreasing diameter which is in qualitative agreement with the theoretical calculations. We conclude that this behavior is not a morphological effect but connected to the properties of the constituting nanotubes and is related to the curvature.

The modified P3 and Laser samples possess the same diameter distribution as their unmodified counterparts, but due to the received acid treatment the constituting nanotubes are doped and their Fermi level moved into one of the Van Hove singularities where the free carrier behavior is not affected by the curvature. Based on this picture we expect that the $E_g$ values of the doped samples fall to zero. The possible explanation of the nonzero $E_g$ is the limited sensitivity of the spectrometer in this low frequency region. These samples exhibit high reflectivity in the far infrared due to increased free carrier concentration. This means low transmission, which makes the measurement ambiguous and easily affected by the instrument's systematic error, complemented by increased error propagation in the Kramers-Kronig transformation near T=0. Nevertheless the observed downshift due to doping is significant and the given explanation is plausible.

\begin{figure}
\includegraphics[width=8 cm]{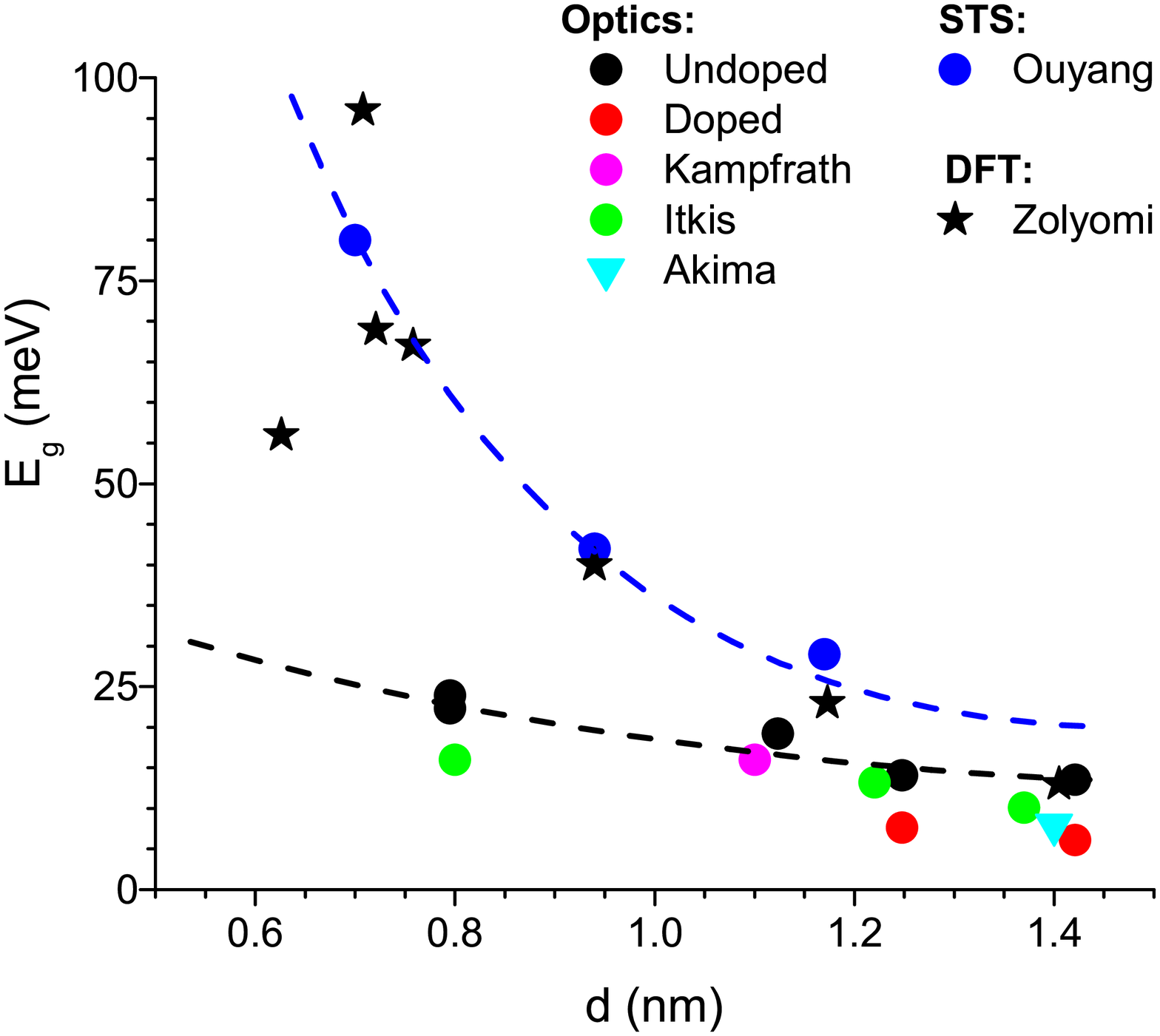}
\caption{\label{fig:compare}  (Color online) Comparison of gap positions determined by various methods: "undoped" and "doped", this study (Fig.~\ref{fig:EgE11}), Kampfrath: photoinduced THz absorption,\cite{kampfrath08}  Itkis: optical absorption,\cite{itkis02} Akima: optical absorption,\cite{akima06} Ouyang: STS,\cite{ouyang01} Z\'olyomi: DFT calculations.\cite{zolyomi04} Dashed lines are guides to the eye.}
\end{figure}

In Fig.~\ref{fig:compare} we compare our results to those of other measurements and to DFT calculations by Z\'olyomi and K\"urti.\cite{zolyomi04} The samples chosen for comparison were commercial materials similar to the ones applied in the present study. We find very good agreement with previously measured optical data, even though the gap values are not strictly comparable due to different evaluation methods. Nevertheless, it is striking that the scatter in the data obtained by optical measurements is minuscule in comparison with the difference between optical and tunneling results, especially for low diameter. Calculated values agree with the tunneling data.

Tunneling is measured on individual nanotubes and therefore the effect of bundling and the environment is less critical; thus it is understandable that these data agree more with theoretical values as those are also obtained for specific $(n,m)$ tubes. Bundling can induce a "pseudogap", but this was predicted to lie way above the curvature gap in frequency\cite{delaney98} and has indeed been observed in STS measurements to be above, not below, the curvature-induced gap;\cite{ouyang01} in inhomogeneous samples, though, this gap is predicted to disappear.\cite{maarouf00} Bundling can therefore be excluded as the reason behind the reduced optical gap values.

Another possibility is connected to the mechanism of the two methods. STS measures the current through the sample and therefore requires extended bands; optical transitions, on the other hand, can occur between localized states as well. Exciton binding energies for the first and second semiconducting transitions have been measured this way.\cite{lin10} Thus, we regard the discrepancy at low diameters a sign of electron-hole interactions even at these small gap values. Further theoretical work is required to predict precise exciton binding energies.

 \section{\label{sec:concl}Conclusion}

We have shown by transmission spectroscopy that the optical conductivity of every nanotube sample exhibits a low-frequency peak. Transmission spectroscopy on solid films (supplemented by Raman and photoluminescence results on suspended nanotubes) can be efficiently used to analyze and assign the optical transitions in macroscopic nanotube samples. The most abundant nanotube species and their average diameter can be defined. The diameter dependence of the low frequency peak is in qualitative agreement with theoretical calculations of the curvature gap. These results indicate that the peak reflects the electronic structure of the nanotubes and not their morphology. Comparing our data with previously measured ones, we find a clear difference between optical gap values on one hand, and tunneling and DFT values on the other, especially for small diameters. This difference invokes the possibility of excitonic effects even in these small gaps.

\begin{acknowledgments}
We thank Jen\H o K\"urti for enlightening comments and discussions.
The research leading to these results has received funding from the European Community's Seventh Framework Programme (FP7/2007-2013) under grant agreement No. 215399 and from the Hungarian National Research Fund (OTKA) under grant No. 75813.
\end{acknowledgments}

%

\end{document}